\documentclass[aps,twocolumn,nofootinbib,prd,aps,10pt]{revtex4}
\usepackage[dvips]{graphicx}
\usepackage[english]{babel}
\usepackage[T1]{fontenc}
\usepackage{mathrsfs}
\usepackage[tbtags]{amsmath}
\usepackage{amssymb}
\usepackage{amsxtra}
\usepackage{amsopn}
\usepackage{latexsym}
\usepackage[mathcal]{eucal}
\usepackage{mathtools}

\newcommand{\BE}{\begin{equation}}
\newcommand{\EE}{\end{equation}}
\newcommand{\BA}{\begin{align}}
\newcommand{\EA}{\end{align}}

\begin{document}

\title{Yang-Mills ghost propagator in linear covariant gauges}

\author{Fabio Siringo}

\affiliation{Dipartimento di Fisica e Astronomia 
dell'Universit\`a di Catania,\\ 
INFN Sezione di Catania,
Via S.Sofia 64, I-95123 Catania, Italy}

\date{\today}

\begin{abstract}
From first principles, using a screened expansion, a simple one-loop analytical expression is provided for the ghost propagator 
of pure SU(3) Yang-Mills theory in a generic linear covariant gauge. At variance with the
Landau gauge, the ghost dressing function is suppressed in the infrared and vanishes at $p=0$, as predicted by other approaches in the
continuum. However, in the very limited range where lattice data are available no detectable deviation is found from the Landau gauge,
thus reconciling some recent lattice data and previous continuum predictions.
\end{abstract}




\maketitle

Usually, in the IR, the gauge-fixed correlation functions of QCD and Yang-Mills theory are studied in the Landau gauge.
The gauge-fixed Faddeev-Popov Lagrangian, which is the most direct way to quantize the theory in perturbation theory, is not
easily implemented in lattice calculations except than for the Landau gauge where the gauge parameter $\xi=0$. 
Even in the continuum, because of the breakdown of the standard perturbation theory, most of the available predictions rely on
numerical calculations at $\xi=0$.

While a qualitative\cite{capri15,papa15,huber15,cucchieri} and even quantitative\cite{bicudo15,xigauge} agreement has been reached
for the gluon propagator at $\xi>0$,
the behavior of the ghost propagator $G(p)$ has not been clearly established yet.
There are strong analytical arguments for a suppression of the ghost dressing function $p^2 G(p)$ in the IR\cite{papa15,huber15,capri16} but there
is no quantitative information on the entity of the suppression and on the scale at which it should be observed.
On the other hand, recent lattice calculations\cite{cucchieri18} indicate that the ghost propagator is
basically gauge parameter independent in the IR, at least at the energies and $\xi$ values explored so far. However, while that
lattice approach to the Faddeev-Popov matrix seems to be very promising, the explored range might be too limited for any final conclusion
to be drawn\cite{cucchieri18}. 

In this paper, it is shown that the recent lattice data of Ref.\cite{cucchieri18} are perfectly consistent with the outcome of
the screened expansion\cite{ptqcd,ptqcd2,analyt,scaling,beta}
which yet predicts the vanishing of the ghost dressing function at $p=0$, as found in Ref.\cite{papa15}. 
From first principles, at one loop, an analytical expression for the ghost dressing function of pure Yang-Mills theory is obtained
by the screened expansion in a generic linear covariant gauge, showing an evident suppression of the dressing function for small momenta $p<0.5$~GeV 
and large values of the gauge parameter $\xi>0.1$, in qualitative agreement with previous calculations in the continuum\cite{huber15,papa15}.
Thus, the apparently contrasting predictions of lattice calculations and continuum approaches are reconciled by the present study, since the
suppression of the ghost dressing function is only observed very deep in the infrared, while no detectable deviation from the Landau gauge is
found in the very limited range where the lattice data are available.

The screened expansion has been proposed in the last years\cite{ptqcd,ptqcd2} and shown to be a powerful tool for describing the 
analytic properties of the correlators of QCD in the IR\cite{analyt,xigauge}. The expansion arises from a change of the expansion point
of ordinary perturbation theory, with the massless free gluon propagator which is replaced by a screened massive propagator
without changing the total action,
providing explicit one-loop analytical expressions which are in excellent agreement with the lattice data\cite{xigauge,scaling,beta}.
Having not modified the total action, the method does not contain any phenomenological parameter and is based on first principles\cite{beta}.
Moreover, its extension to finite temperature\cite{damp,varT} and to a generic covariant gauge\cite{xigauge} has allowed the study of several
features, like the gauge invariance and temperature dependence of the gluon dispersion relations, which cannot be observed by lattice calculations.
In fact, like for ordinary perturbation theory, the extension to a generic linear covariant gauge is straightforward and if the expansion is
optimized by the constraints of the exact Becchi-Rouet-Stora-Tyutin (BRST) symmetry of the action, the gluon propagator can be evaluated for any value of the gauge parameter
$\xi$, yielding reliable quantitative predictions even in the Feynman gauge which is not too much explored yet\cite{varqcd,genself,highord}.
For the gluon propagator, the agreement with the lattice is very good in the limited range $\xi<0.5$ where the data are available\cite{bicudo15}.
Thus, it would be now interesting to compare the recent data of Ref.\cite{cucchieri18} with the outcome of the screened expansion for the ghost propagator.

The ghost sector and the running coupling have been already studied by the screened expansion in a very recent work\cite{beta} 
where the general renormalization of the expansion
is discussed in the Landau gauge and some general stationary conditions are derived for the optimization of the finite parts of the renormalization constants,
from first principles. The analytical predictions of the expansion are found in excellent agreement with the lattice data of Ref.\cite{duarte}
for the correlators and for the strong coupling. 

\begin{figure}[b] \label{fig:dressing}
\hspace*{-1cm}\includegraphics[width=0.4\textwidth,angle=-90]{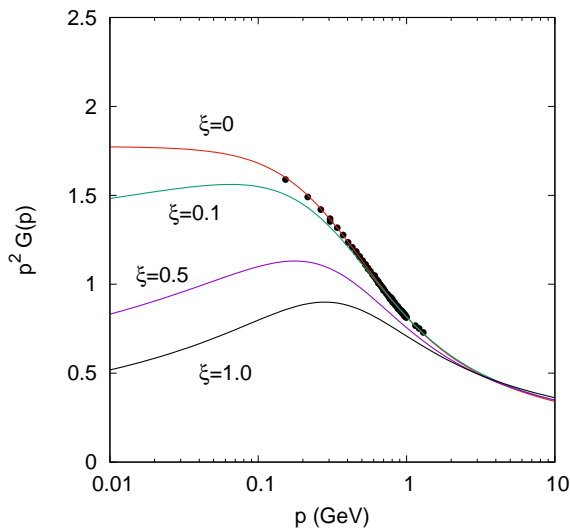}
\caption{The SU(3) ghost dressing function $p^2 G(p)$ is evaluated by Eq.(\ref{dressing}) for different values of $\xi$, ranging from the Landau ($\xi=0$)
to the Feynman gauge ($\xi=1$), and plotted together with some lattice  data of Ref.\cite{duarte} in the Landau gauge ($\beta=6$, $L=80$).}
\end{figure}

\begin{figure}[b] \label{fig:propagator}
\includegraphics[width=0.35\textwidth,angle=-90]{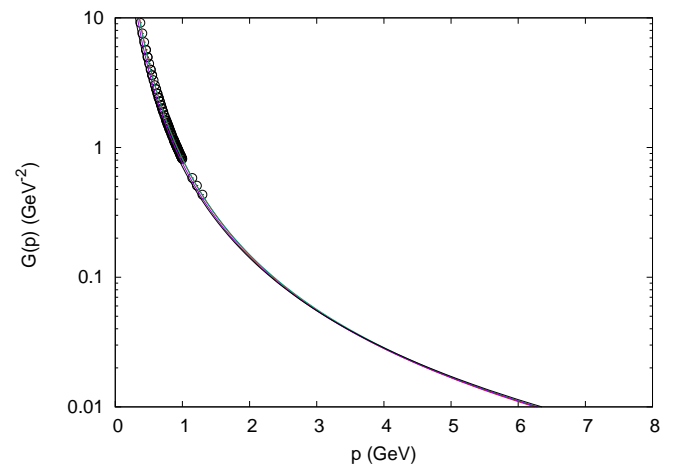}
\caption{The SU(3) ghost propagator $G(p)$ is evalauated by Eq.(\ref{dressing}) and shown for the same values of the gauge parameter as in Fig.~1, 
together with the same lattice data in the Landau gauge, but on the log scale used in Ref.\cite{cucchieri18}. 
Observe the exchange of the log and linear scales, between
the two axes, compared to Fig.~1.}
\end{figure}

While that work was in the Landau gauge, the extension to a generic covariant gauge is trivial for the
ghost propagator at one-loop, since the gauge parameter enters through the longitudinal part of the gluon propagator of a unique one-loop graph\cite{xigauge}.
The longitudinal part of the gluon propagator is known exactly and is equal to the free longitudinal part of ordinary perturbation theory. Thus, for
a generic $\xi\not=0$, the new self-energy contribution is just the same as that of perturbation theory and by dimensional regularization, for $d=4-\epsilon$, it reads
\BE
\Sigma_\xi(p)=\xi\,\frac{N}{4}\left(\frac{\alpha_s}{4\pi}\right)\, p^2 \left(\frac{2}{\epsilon}+\ln\frac{\mu^2}{p^2} + C\right)
\EE
where $\alpha_s=g^2/(4\pi)$ is the strong coupling, $\mu$ is an arbitrary energy scale, $C$ is an arbitrary constant and $p^2$ is the momentum in the
Euclidean formalism.
The diverging part gets canceled by the usual counterterm arising from wave function renormalization, so that 
the finite part of the total ghost self energy can be written as
\BE
\Sigma(p)=3N\left(\frac{\alpha_s}{4\pi}\right)\, p^2 \left[{\cal G}(s)-\frac{\xi}{12}\ln s\right]
\EE
where $s=p^2/m^2(\xi)$ and $m(\xi)$ is a gauge-dependent mass parameter which was determined in Ref.\cite{xigauge} by requiring that
the poles of the gluon propagator are gauge parameter independent, as imposed by Nielsen identities. A very accurate interpolation of
the curve $m(\xi)$ in the range $0<\xi<1$ was determined in\cite{xigauge}
\BE
m^2(\xi)/m_0^2\approx 1-0.39997\,\xi+0.064141\,\xi^2
\EE
where the mass parameter $m_0$ must be regarded as a phenomenological energy scale which fixes the units.
By a comparison with the lattice units of Ref.\cite{duarte}, the value $m_0=0.656$ GeV was extracted\cite{xigauge}.
The adimensional function ${\cal G}(s)$ gives the finite part of the self energy in the Landau gauge and its explicit analytical expression has been reported
in several papers\cite{ptqcd,ptqcd2,analyt,beta} 
\begin{align}
{\cal G}(s)&=
\frac{(1+s)^2(2s-1)}{12 s^2}\ln(1+s)-\frac{s}{6}\ln(s)+\frac{1+2s}{12s}.
\end{align}

Following the scheme of Ref.\cite{beta}, up to a finite renormalization factor,
the ghost dressing function reads
\BE
p^2 G(p)=\left[{\cal G}(s)-\frac{\xi}{12}\ln s+{\cal G}_0\right]^{-1}
\label{dressing}
\EE
where the constant ${\cal G}_0$ depends on the renormalization scheme and can be optimized by the stationary conditions
of Ref.\cite{beta}, yielding ${\cal G}_0=0.14524$ at $\xi=0$ for SU(3). Using the same  conditions, it can be checked that, for a generic gauge, the variation
of the optimal constant ${\cal G}_0$ is small and of order
$\delta {\cal G}_0/{\cal G}_0\approx 0.1\,\xi$, so that its dependence on $\xi$ can be safely neglected at the scale of the figures shown below.

By inspection of Eq.(\ref{dressing}), we observe that the dressing function is logarithmically suppressed in the IR for any $\xi\not=0$. 
Moreover, since ${\cal G}(s)$
is finite in the limit ${s\to 0}$~\cite{beta}, the dressing function is zero at $p=0$, as predicted in Ref.\cite{papa15}.
In that work, because of the self consistency of the equations, the leading behavior was $(\xi \ln p)^{-1/2}$ while here, 
the simple one-loop dressing function vanishes at the stronger rate $(\xi \ln p)^{-1}$. Thus we expect that in Eq.(\ref{dressing})
the effect might be overestimated deep in the IR.

The SU(3) dressing function is shown in Fig.~1 for several values of the gauge parameter $\xi$ ranging from the Landau to the Feynman gauge.
In the figure, the curves have an arbitrary renormalization at the scale $\mu= 4$~GeV. In the same figure, some large-volume lattice data of Ref.\cite{duarte} are
shown for the Landau gauge. The data have been renormalized in order to match the corresponding curve for $p< 2$ GeV. As discussed in Ref.\cite{beta} no
comparison can be made above that point because the fixed-coupling one-loop dressing function needs to be corrected by RG effects. 

The dressing function
has a maximum for any $\xi\not=0$, but the location of the maximum is at the very small momentum $p=0.064$ GeV for $\xi=0.1$, which is the largest
$\xi$ value reached in Ref.\cite{cucchieri18}. The maximum moves to $p=0.18$ GeV and $p=0.28$ GeV for $\xi=0.5$ and $\xi=1$, respectively.
Even in the Feynman gauge, the maximum is far below $p=0.5$ GeV which is the smaller momentum in Ref.\cite{cucchieri18}. 

Above $p=0.5$, the curves
are not distinguishable for $\xi\le 0.1$, in perfect agreement with the preliminary results of Ref.\cite{cucchieri18} which show no 
detectable difference with the data in the Landau gauge. However, Fig.~1 shows that a difference should be observable in the lattice data if a smaller
momentum $p\approx 0.2$ GeV is reached, or even at $p=0.5$ GeV if a larger gauge parameter is taken (say $\xi=0.5$ at least).

We must mention that the effect in Fig.~1 has been magnified by plotting the dressing function $p^2 G(p)$ 
on a linear scale instead of the diverging propagator $G(p)$ which was shown on a log scale in  Ref.\cite{cucchieri18}. Actually, as shown in
Fig.~2, if the propagator is plotted on the same scale of Ref.\cite{cucchieri18}, the differences are not detectable any more and even in the
Feynman gauge the curve follows the data of the Landau gauge. Thus, no conclusion can be drawn from the preliminary lattice data of 
Ref.\cite{cucchieri18} unless the authors can manage to show the dressing function, as in Fig.~1.
It would be a very important test for the predictions of the screened expansion if still no difference could be observed in
the data on that scale. Of course, if smaller energies or larger gauge parameters could be reached on the lattice by that new method,
then a more complete test of the expansion would be available.

In summary, we have shown that the screened expansion provides a very simple one-loop expression for the ghost propagator in
a linear covariant gauge. The SU(3) dressing function and propagator have been shown for several values of the gauge parameter $\xi$,
up to the Feynman gauge, and compared with the very recent lattice data of Ref.\cite{cucchieri18}. 
Even if the data seem to suggest that the ghost propagator is gauge parameter independent, we argue that no final
conclusion can be drawn from the data because of the limited explored range of $\xi$ and $p$. In fact, while the analytical curves are in perfect agreement
with the data in that range, they show an evident suppression of the ghost dressing function for smaller momenta $p<0.5$~GeV and larger values
of the parameter $\xi>0.1$, in qualitative agreement with previous calculations in the continuum\cite{huber15,papa15}.
Moreover, we point out that in order to amplify the effect and make it observable, the ghost dressing function should be plotted on a linear scale
rather than the diverging propagator on a log scale. Otherwise, as shown in Fig.~2, no effect would be seen since the propagator still diverges even for
$\xi\not =0$, hiding the effect of the suppression in the IR. More extended data, by the method of Ref.\cite{cucchieri18}, will be very welcome
for providing a more stringent comparison with the present and previous analytical predictions.

The author acknowledges a partial financial support from "Piano per la Ricerca di Ateneo 2017/2020 -
Linea di intervento 2" of the University of Catania.


\begin{thebibliography} {99}


\bibitem{capri15} M. A. L. Capri, A. D. Pereira, R. F. Sobreiro, S. P. Sorella, Eur. Phys. J. C {\bf 75}, 479 (2015).
\bibitem{papa15} A. C. Aguilar, D. Binosi, J. Papavassiliou, Phys. Rev. D {\bf 91}, 085014 (2015).
\bibitem{huber15} M. Q. Huber, Phys. Rev. D 91, 085018 (2015).
\bibitem{cucchieri} A. Cucchieri, T. Mendes, E. M. S. Santos, Phys. Rev. Lett. {\bf 103}, 141602 (2009). 
\bibitem{bicudo15} P. Bicudo, D. Binosi, N. Cardoso, O. Oliveira, P. J. Silva, Phys. Rev. D {\bf 92}, 114514 (2015).
\bibitem{xigauge} F. Siringo and G. Comitini, Phys Rev. D {\bf  98}, 034023 (2018).
\bibitem{capri16} M. A. L. Capri, D. Fiorentini, M. S. Guimaraes, B.W. Mintz, L. F. Palhares, S. P. Sorella, D. Dudal, I. F. Justo,
A. D. Pereira, R. F. Sobreiro, Phys. Rev. D {\bf 93}, 065019 (2016).
\bibitem{cucchieri18}  A. Cucchieri, D. Dudal, T. Mendes, O. Oliveira, M. Roelfs, P. J. Silva, Phys. Rev. D {\bf 98}, 091504(R) (2018).

\bibitem{ptqcd}  F. Siringo, {\it Perturbative study of Yang-Mills theory in the infrared}, arXiv:1509.05891.
\bibitem{ptqcd2} F. Siringo, Nucl. Phys. B {\bf 907}, 572 (2016), [arXiv:1511.01015].
\bibitem{analyt} F. Siringo, Phys. Rev. D {\bf 94}, 114036 (2016), [arXiv:1605.07357].
\bibitem{scaling} F. Siringo, EPJ Web of Conferences {\bf 137}, 13016 (2017), [arXiv:1607.02040].
\bibitem{beta} F.Siringo, arXiv:1902.04110.

\bibitem{damp}   F. Siringo, Phys. Rev. D {\bf 96}, 114020 (2017), [arXiv:1705.06160].
\bibitem{varT} G. Comitini, F. Siringo, Phys. Rev. D {\bf 97}, 056013 (2018).

\bibitem{varqcd} F. Siringo, Phys. Rev. D {\bf  90}, 094021 (2014). 
\bibitem{genself} F. Siringo, Phys. Rev. D {\bf 92}, 074034 (2015). 
\bibitem{highord} F. Siringo, Phys. Rev. D {\bf 88}, 056020 (2013).
\bibitem{duarte} A. G. Duarte, O. Oliveira, P. J. Silva, Phys. Rev. D {\bf 94}, 014502 (2016).

\end{thebibliography}
\end{document}